# What's in a Text-to-Image Prompt? The Potential of Stable Diffusion in Visual Arts Education

Nassim Dehouche [1,*], Kullathida Dehouche [2]

[1] Business Administration Division, Mahidol University International College, Salaya, Thailand; nassim.deh@mahidol.edu
[2] Poh-Chang Academy of Arts, Rajamangala University of Technology Rattanakosin, Bangkok, Thailand; kullathida.mee@rmutr.ac.th
[*] Correspondence: nassim.deh@mahidol.edu; Mahidol University International College, 999 Phuttamonthon 4 Road, Salaya. 73170, Thailand

**Abstract:** Text-to-Image artificial intelligence (AI) recently saw a major breakthrough with the release of Dall-E and its open-source counterpart, Stable Diffusion. These programs allow anyone to create original visual art pieces by simply providing descriptions in natural language (prompts). Using a sample of 72,980 Stable Diffusion prompts, we propose a formalization of this new medium of art creation and assess its potential for teaching the history of art, aesthetics, and technique. Our findings indicate that text-to-Image AI has the potential to revolutionize the way art is taught, offering new, cost-effective possibilities for experimentation and expression. However, it also raises important questions about the ownership of artistic works. As more and more art is created using these programs, it will be crucial to establish new legal and economic models to protect the rights of artists.

**Keywords:** artificial intelligence; art; education; computational creativity; intellectual property.

> "It is, in the first place, 'by a word conceived in intellect' that the artist, whether human or divine, works." Ananda K. Coomaraswamy [1]

## 1. Introduction

The traditional view of art, espoused by Coomaraswamy [1], is that of (human) art as imitation (of divine creation), with the word as a starting point. This view, notably challenged by contemporary expressionist and formalist perspectives [2], was given a new technical expression with recent advances in artificial intelligence (AI).

Indeed, AI has made impressive strides in the realm of creativity, with computers now able to generate relevant and original text [3] and images [4, 5], in response to simple natural language prompts. Some of these outputs have even been indistinguishable from human creations, leading to their recognition in traditional art contests [6].

AI-generated art remains a controversial topic, with notable debates over whether it can truly be considered art in the first place [7], but despite the increasing academic interest in generative AI models, little attention has been given to their potential use in visual arts education. In our view, these models contain a compressed version of centuries of human artistic creations, which presents an undeniable interest for art education. Thus, in this paper, we explore the possibilities of incorporating them in visual art education, particularly for the teaching of art history, aesthetics, and technique.

Following this introductory section, the remainder of this paper is organized as follows. Section 2 situates recent developments in the field of Text-to-Image in the broader history of AI-generated art. Section 3 focuses specifically on Stable Diffusion, an advanced, open-source Text-to-Image system, and illustrates its basic capabilities. Section 4 describes the methods and data of our analysis of 72,980 Stable Diffusion interactions. Based on this analysis, Section 5 proposes a formalization and procedural framework for Stable Diffusion prompts that can serve as a basis for their formal usage in educational software or curricula, and discusses some of its potential uses for the teaching of subjects such as the history of art, aesthetics, and technique, as well as its implications for the protection of the intellectual property of artists. Lastly, Section 6 concludes this paper by outlining the work that remains to be done, in our view, to facilitate the integration of Text-to-Image AI in art education.

## 2. A Brief History of AI-generated Art

The first attempts at using Artificial Intelligence to create coherent, original content from human prompts can be traced back to the 1950s, when researchers at the MIT Artificial Intelligence Laboratory created a program called ELIZA [8]. ELIZA was able to generate simple responses to text input, using pattern matching and natural language processing techniques. While not strictly art, ELIZA was an early example of Text-to-Text: software that could generate original text output that was intended to be interpreted by humans. One of the first examples of AI-generated art proper was a program called AARON, developed by artist Harold Cohen in the 1970s [9]. AARON was a computer program that was capable of generating complex drawings and paintings. AARON used a set of rules and constraints to create its art, and was able to learn from its own outputs to improve over time.

As AI technology advanced in the 1980s and 1990s, more complex and sophisticated AI-generated art began to emerge. For instance, Karl Sims generated unique 3D images and animations based on evolutionary algorithms [10]. In recent years, the advent of deep learning has led to even more realistic outputs, and consequently, AI-generated art gained increasing attention from both the art world and the general public. In 2015, a team at Google used deep learning techniques to train a neural network on a dataset of over 10,000 paintings, with the goal of generating original works of art from input images. The resulting program, known as DeepDream [11], was able to create surreal, visually striking images from input images (Image-to-Image). Another notable example is the work of a Paris-based art collective named "Obvious," which resulted in a software-generated portrait that sold for over $432,000 at a Christie's auction, in 2018 [12].

Year 2020 saw a major qualitative leap in Text-to-Text capabilities, with the release of the third generation Generative Pretrained Transformer (GPT-3), by private research firm OpenAI [3]. GPT-3 constitutes an important advance in terms of the generality of Text-to-Text models, and is able to generate text that is highly coherent, in response to virtually any prompt in natural language. This was made possible by the sheer size of the model, which consisted of 175 billion parameters; an order of magnitude more than the second largest similar model to date. This vast number of parameters allowed GPT-3 to comprehend language tasks it was not particularly trained for, and ushered in the era of Large Language Models. These models have the ability to generate high-quality, human-like text, which can be used in a variety of applications, including machine translation, text summarization, and creative writing. The success of GPT-3 led to the development of CLIP [13], another breakthrough model by OpenAI, which was designed to link text to images. CLIP (Contrastive Language–Image Pretraining) is a general-purpose image-text model trained on 400 million text-image pairs from the internet, allowing it to perform image classification with any user-provided label. It can also generate text that accurately describes any input image (Image-to-Text). Based on these advances, OpenAI released DALL-E [4], which is able to generate convincing images from text descriptions (Text-to-Image). While DALL-E remains a proprietary, closed-source software, the code of CLIP was released open-source. This allowed artificial intelligence firm Stability AI to develop and train Stable Diffusion [5], an open-source Text-to-Image model, with comparable performance to DALL-E. Stable Diffusion was released under a permissive license allowing commercial and non-commercial usage.

Although they represent an important technical breakthrough, CLIP, and the Text-to-Image systems based on it, also raise important ethical and societal concerns. Because of its training on mass, indiscriminate internet data, CLIP has a propensity to reproduce biased and unfair stereotypes present in culture and society [14], and its possible unfair usage of protected works has alerted legal experts [15]. These systems also have the potential to be used for nefarious purposes, such as creating fake news or spreading misinformation [16].

**3. Stable Diffusion**

Stable Diffusion is a text-to-image model, released in 2022, that uses a deep learning technique called latent diffusion [5] to generate images based on text descriptions. Unlike some previous Text-to-Image models, Stable Diffusion's code and model weights are publicly available and can be run on most consumer hardware.

To generate images, Stable Diffusion uses CLIP [12] to project a text prompt into a joint text-image embedding space, and select a rough, noisy image that is semantically close to the input prompt. This image is then subject to a denoising method based on the latent diffusion model to produce the final image. In addition to a text prompt, the Text-to-Image generation script within Stable Diffusion allows users to input various parameters such as sampling type, output image dimensions, and seed value.

This latter integer parameter is typically set randomly, but a constant seed value allows for reproducibility, and the conservation of some aspects of the generated images across prompts. For instance, in Figure 1(a). and Figure 1(b)., we use Stable Diffusion version 2.1 to generate two images, with the respective prompts "detailed photograph of an older woman/man wearing a leather jacket, waist shot, forest background, in the style of Brandon Stanton, Humans of New York", and set the seed in both images at a value 1034. We can see that this seed value conserves some of the facial features of the subject, across prompts and genders. Additionally, a constant seed value can be useful to maintain a subject's appearance in different poses and settings. For instance, Figure 2. shows images resulting from the prompt "digital illustration of an older woman/man wearing a leather jacket, Victorian aesthetics, waist shot, forest background, in the style of Magali Villeneuve" and a constant seed value of 1242. It should be noted that, even for a constant seed value, text prompts typically generate some random artifacts and imperfection, such as the variations in the neckwear of the character between Figure 2(a) and 2(b), as well as Figure 2(c) and 2(d). These imperfections can require additional post-processing of the generated images.

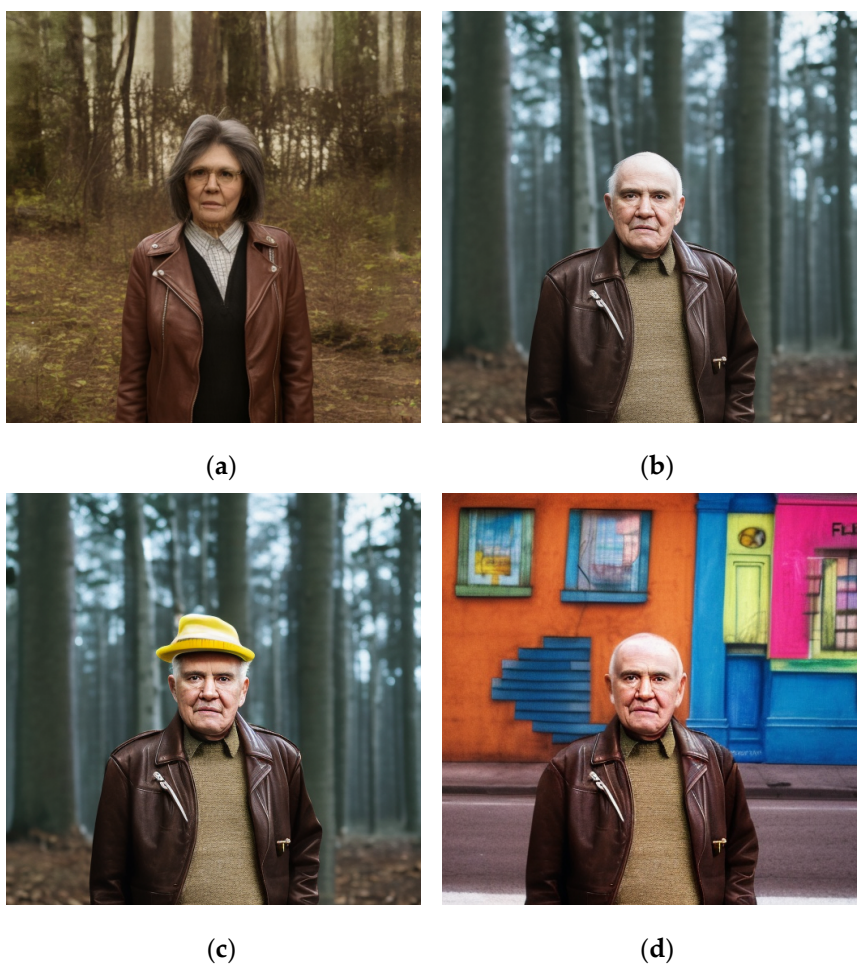

**Figure 1.** Images generated in Stable Diffusion 2.1., with the prompts "detailed photograph of an older woman/man wearing a leather jacket, waist shot, forest background, in the style of Brandon Stanton, Humans of New York". Additional inpainting was applied to generate Figures (c) and (d).

Some front-end implementations of Stable Diffusion, such as DreamStudio[1], offer additional functions for post-processing tasks, such as inpainting and outpainting. Inpainting involves altering a specific part of an image by filling in a masked area with new content based on a user-provided prompt. Outpainting, on the other hand, involves generating new content to extend an image beyond its original dimensions based on a user-provided prompt. Both of these functions use the Stable Diffusion model to generate the new content. For instance, with Figure 1(b) as a starting point, we can add accessories to the subject, as illustrated in Figure 1(c), or change the background of the image, as in Figure 1(d), with the respective inpainting prompts "man wearing a yellow hat" and "man in a colorful street corner".

---

[1] https://beta.dreamstudio.ai/

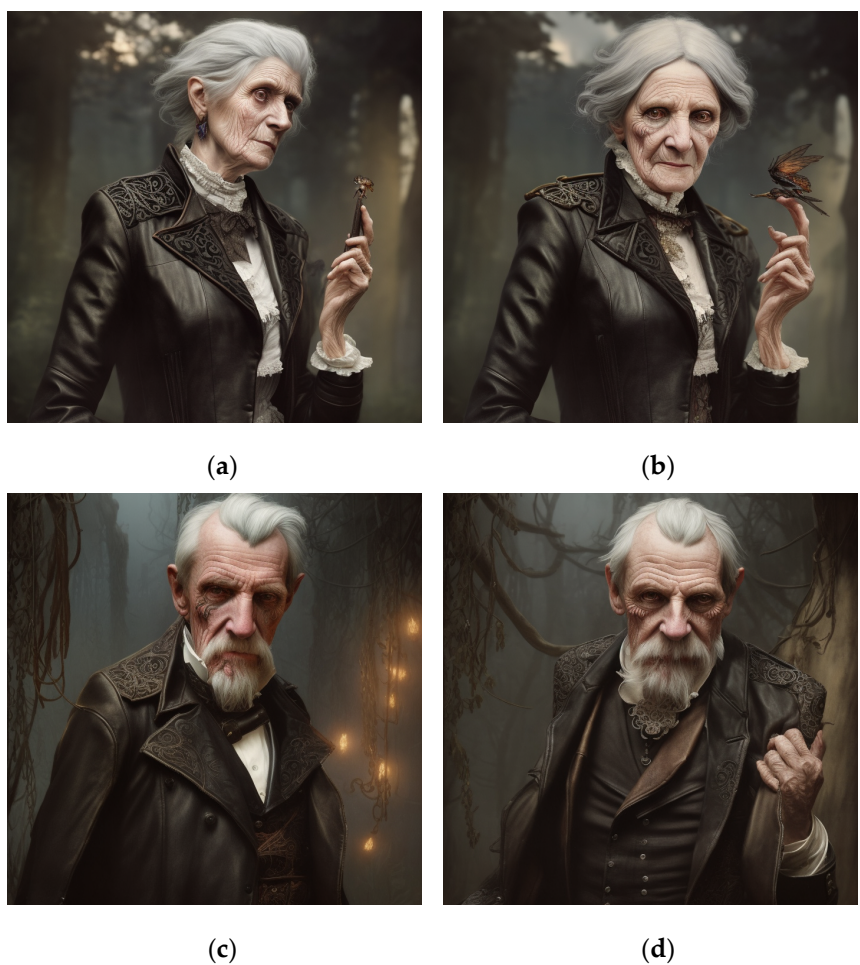

(**a**) (**b**)

(**c**) (**d**)

**Figure 2.** Images generated in Stable Diffusion 2.1., with the prompts "digital illustration of an older woman/man wearing a leather jacket, Victorian aesthetics, waist shot, forest background, in the style of Magali Villeneuve".

Moreover, Figure 1. and Figure 2. illustrate Stable Diffusion's ability to reproduce the style of contemporary, practicing artists (photographer Brandon Stanton and illustrator Magali Villeneuve, respectively). This controversial aspect of generative AI [17] is analyzed more thoroughly in Section 4.3.

### 4. Data and Methods

Stable Diffusion's output images are highly sensitive to the wording of text prompts, so we set out to examine the format and semantic content of this form of input. To this end, we gathered a dataset of 72,980 Stable Diffusion prompts from Lexica[2], a search engine that features curated Stable Diffusion outputs submitted by users along with the prompts that generated them. We conducted our analysis in three steps:

1. Tokenization: Each prompt is broken down into "tokens"; atomic linguistic terms, which can be words, phrases, symbols, or other meaningful elements of the prompt. This step is performed using the BERT Tokenizer [18].
2. Topic extraction: The goal of this step is to automatically identify the main topics or themes present in the 72,980 prompts, with the prior knowledge that they represent detailed description of images. This is performed using the GPT-3 [3] API[3].

---

[2] https://lexica.art/
[3] https://openai.com/api/

3. Classification: Tokens, from each prompt, are classified into one or several of the linguistic topics identified in step 1, using the GPT-3 API.

Additionally, the ability of Stable Diffusion to accurately reproduce the style of specific artists, whose work was used for its training, has been a controversial issue. To specifically examine the usage of this feature in prompts, we identified tokens that represent the name of an artist, brand, or collective using BERT's named-entity recognition function [18] and calculated the frequency of each of these entities in the 72,980 prompts under consideration.

## 5. Results and Discussion

### 5.1. Formalizing Stable Diffusion Prompts

Topic extraction allows us to identify the primary elements (i.e. semantic categories of tokens) described in Table 1. These are the most frequent categories of keywords in the 72,980 considered prompts.

**Table 1.** Primary elements in 72,980 Stable Diffusion prompts

| Topic | Description |
|---|---|
| Subject | The characters and objects in the image, such as "a cyborg", "two dogs", "a car", "a wizard", etc. |
| Medium | The type of visual object that is the image, such as "digital illustration", "photograph", "3D render", "concept art", "poster", etc. |
| Technique | The tools and software used to create the image, such as "Blender", "pincushion lens", "Unreal engine", "Octane", etc. |
| Genre | Aesthetic features that describe the artistic genre of the image, such as "anime", "surreal", "baroque", "photorealistic", sci-fi, black and white, epic fantasy, film noir, etc. |
| Mood | Features that describe the atmosphere and emotions of the image, such as "beautiful", "eerie", "bleak", etc. |
| Tone | Features that describe the chromatic composition of the image, such as "pastel", "synthwave colors", "ethereal colors", etc. |
| Lighting | The use of light and shadows in the image "dark", "cinematic lighting", "realistic shaded lighting", "studio lighting", radiant light, etc. |
| Resolution | Features that describe the level of detail of the image, e.g. "highly-detailed", "photorealistic", "100 mm", "8K", "16K", "HQ", "sharp focus", etc. |
| Artistic References | Artists or works of art to use as inspiration, e.g. "Greg Rutkowski", "Studio Ghibli", "Artgerm", "Zaha Hadid", etc. |
| Reception/Popularity | Awards, recognition, or trending status on art-focused platforms,. e.g. "trending on artstation", "masterpiece", "award-winning", etc. |

Less frequent topics, that are extensions or additional details of the previous main topics are listed in Table 2.

**Table 2.** Secondary elements in 72,980 Stable Diffusion prompts

| Topic | Examples |
|---|---|
| Physical attributes of the subject | race, age, clothing, accessories, "cute", "glamorous", "chonky", etc. |
| Emotional or psychological traits of the subject | "happy", "anxious", "triumphant", "pensive", etc. |
| Environment/Setting | time, weather, "medieval", "post-apocalyptic", etc. |
| Symmetry/Repetition | "symmetry", "symmetrical", "pattern", "motif", "fractal", etc. |
| Depth of field | "blurred background", "deep focus", "aperture", "F/4", "F/2.8", "sharp focus", "bokeh", etc. |
| Angle | "ultra wide angle", "zenith view", "cinematic view", "close up", etc. |
| Message/Meaning | "propaganda", "religious', "advertisement", etc. |

### 5.2. Proposed Procedural Classification

The identified prompt elements align remarkably well with traditional photography concepts, and can be procedurally classified as in Figure 3.

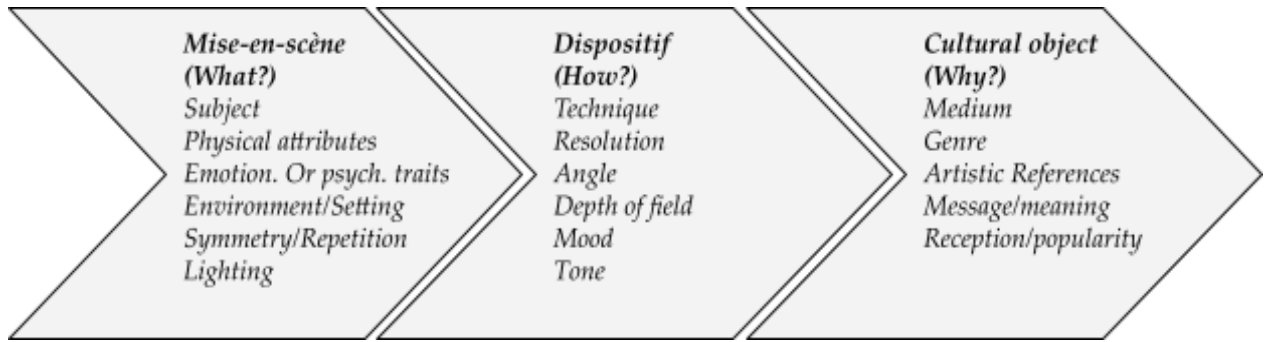

**Figure 3.** Proposed creative process for Text-to-Image prompts based on the semantic elements in 72,980 Stable Diffusion prompts.

- **M*ise-en-scène*:** Mise-en-scène is a term commonly used in the study of photography, film, and theater to refer to the arrangement of objects, settings, and actors within a shot or scene [20]. This category thus includes the visual and compositional elements that will appear in the frame to create the intended cultural object, e.g. "a defiant older woman/man wearing a leather jacket, in a post-apocalyptic city, bleak lighting", illustrated in Figure 4.

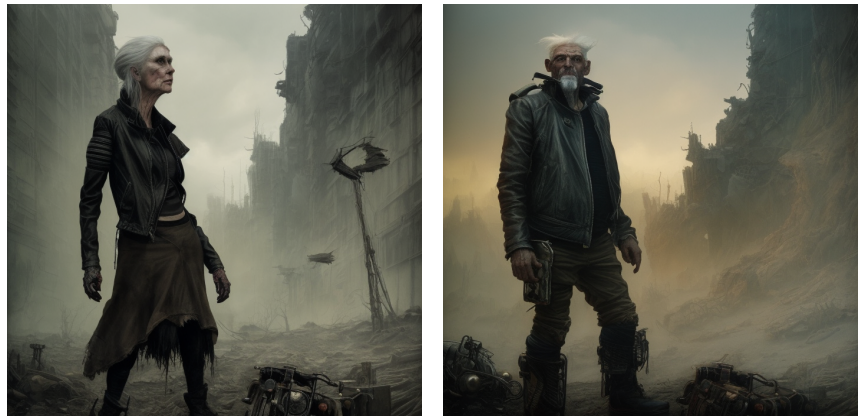

(**a**) (**b**)

**Figure 4.** Images generated in Stable Diffusion 2.1., with the prompts "digital painting of a defiant older woman/man wearing a leather jacket, in a post-apocalyptic city, bleak lighting, trending on artstation, Greg Rutkowski"

- *Dispositif:* In photography and film, the concept of dispositif pertains to the configuration of the material technology [19] used to capture an image. Within our more general classification, this category can also possibly include software tools and post-processing techniques for digital images. If mise-en-scène is *what* is displayed in the image, the dispositif would be *how* it is created, e.g. "close up, black and white, wide aperture, 8K, sharp edges", illustrated in Figure 5.

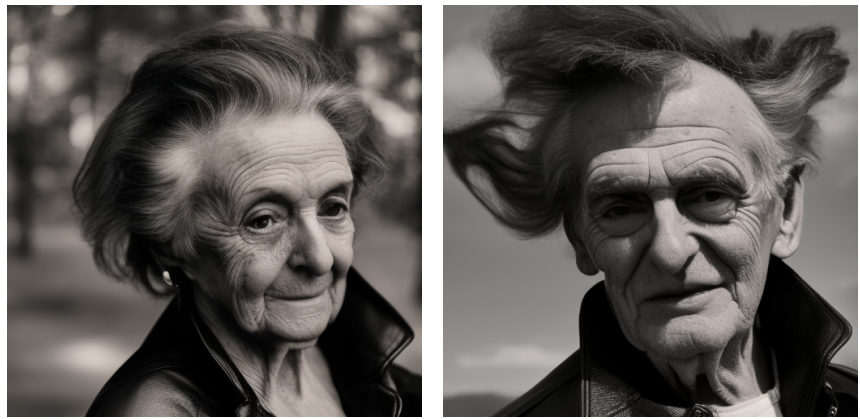

(**a**) (**b**)

**Figure 5.** Images generated in Stable Diffusion 2.1., with the prompts "portrait photograph of a happy, pensive older woman/man wearing a leather jacket, forest background, close up, black and white, wide aperture, 8K, sharp edges, Robert Doisneau".

- *Cultural object:* These elements describe the "object" of the artist's creation, understood in its double meaning of "artifact" and "purpose"; the latter understanding includes descriptions of the medium and genre of the image, as well as its positioning in the history of art through artistic references (e.g. "a photograph by Annie Leibovitz" or "a renaissance painting by Michelangelo"); the former descriptions of the message/meaning and reception/popularity (e.g. "religious, award-winning"). These two combinations of prompts are illustrated in Figure 6.

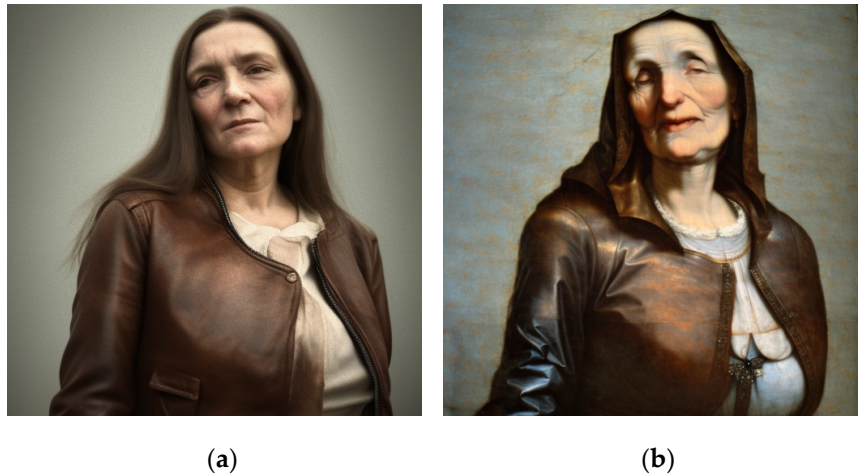

(**a**)            (**b**)

**Figure 6.** "portrait of an older woman wearing a leather jacket, religious, award-winning", as (a) "a photograph by Annie Leibovitz" and (b) "a renaissance painting by Michelangelo".

It is important to note that the elements in our proposed procedural classification are not independent or exclusive. For example, using an artist's name as an artistic reference can influence the mood and tone of the resulting image. It can be interesting to explore unusual or conflicting combinations of these elements for creative purposes, but it is worth remembering that the initial image associated with a text by CLIP is a noisy pixel soup, and the prompts are meant to guide its denoising. Therefore, the more coherent the prompt, the better the outcome. Mastering Text-to-Image involves understanding the interplay of these elements, which includes a degree of randomness, in order to generate the most coherent art.

**5.3. The Need for New Economic Models for Visual Arts**

The word cloud in Figure 7. shows the frequency of named entities used as artistic references in the 72,980 prompts under consideration. We found that these named entities predominantly refer to contemporary, practicing artists who frequently post their work on digital art platform ArtStation. For example, Polish painter Greg Rutkowski, including slight misspellings of his name and mentions alongside other artists, appears in 41.06% of the prompts, while mentions of ArtStation as an element of Reception/Popularity appear in 63.35% of the prompts.

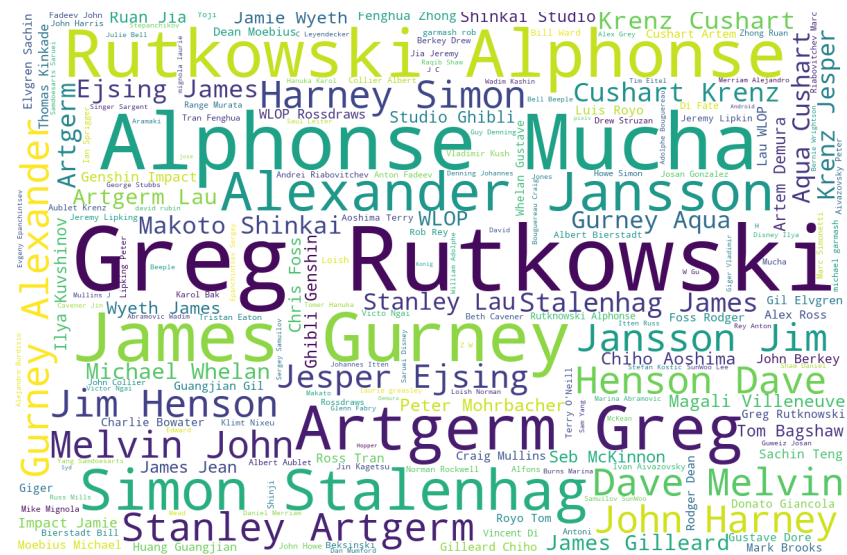

**Figure 7.** Word cloud of artists, brands, or collective names used for inspiration in 72,980 Stable Diffusion prompts.

The popularity of these keywords can be attributed to the fact that platforms like ArtStation encourage artists to include detailed labels describing their work in order to make it more accessible to persons with disabilities, which makes these creations particularly useful for training Text-to-Image AI models. Thus, ArtStation artists are somehow penalized for their virtue. The legal question of whether this training constitutes plagiarism is still open [15] and may take years, if not decades, to be resolved. In addition to possible unfair usage of the intellectual property of these artists, the widespread use of Stable Diffusion also leads to the original creations of these artists being overshadowed in search engine results by AI-generated works that bear their names in the prompts.

While incomplete, as it does not account for works that are used implicitly in the creation of an image, a simple technical solution to these issues could be to devise compensation models for artists based on the frequency of their names appearing as a Style Reference in commercial Text-to-Image applications, similar to music streaming economic models.

## 6. Conclusions

This paper aims to provide a structured approach to understanding this new medium of art creation and connect it to established art education concepts, despite the fact that the output of a Stable Diffusion prompt is random to some extent and highly sensitive to its wording.

Stable Diffusion's "understanding" of art is no doubt superficial and essentially situated at the level of gimmicks (which, as noted by [21], remain "capitalism's most successful aesthetic category"). However, with proper guidance and curation from educators, it can represent a valuable, didactic tool for the transmission of technical concepts, as well as more experiential concepts of artistic genres, movements, and aesthetics that characterize a cultural object. Additionally, variations on the elements of mise-en-scène and dispositif, for a constant seed integer, can constitute a fast and cheap method of experimentation and prototyping, before using costly studio time.

Notwithstanding their potential, for Stable Diffusion and similar software to be harmoniously integrated into the art world, it is necessary for there to be ethical and legal clarity surrounding the important questions they raise about the fair compensation for artists whose creations were used to train these models.